\documentstyle [12pt,epsfig]{article}

\input{epsf}
\begin{document}
\begin{titlepage}
\begin{center}

\vspace{-0.7in}

{\large \bf Vacuum Stress Tensor of a Scalar Field \\in a Rectangular
Waveguide \\}
\vspace{.3in}

{\large\em R.B. $Rodrigues^{a,}$\footnote[1]{email: robson@cbpf.br},
           N.F. $Svaiter^{a,}$\footnote[2]{Present address: Center for      
           Theoretical Physics, Laboratory for Nuclear Physics, Massachussets 
           Institute of Technology; email: svaiter@lns.mit.edu}
       and R.D.M. De $Paola^{b,}$\footnote[3]{email: rpaola@efei.br}}\\
\vspace{.4in}
$^{a}$Centro Brasileiro de Pesquisas Fisicas-CBPF\\ Rua Dr. Xavier 
Sigaud 150, Rio de Janeiro, RJ, 22290-180, Brazil\\
$^{b}$Instituto de Ci\^encias - Escola Federal de Engenharia de Itajub\'a\\
Av. BPS 1303 Pinheirinho, 37500-903, Itajub\'a, MG - Brazil

\subsection*{\\Abstract}
\end{center}

Using the heat kernel method and the analytic continuation of the zeta 
function, we calculate the canonical and improved vacuum 
stress tensors, $\left\langle T_{\mu \nu }(\vec{x})\right\rangle $ and 
$\left\langle \Theta _{\mu \nu }(\vec{x})\right\rangle $, associated with a 
massless scalar field confined in the interior of an infinitely long rectangular
waveguide. The local depence of the renormalized energy for two special 
configurations when the total energy is positive and negative are presented
using $\left\langle T_{00}(\vec{x})\right\rangle $  and 
$\left\langle \Theta _{00 }(\vec{x})\right\rangle $. From the stress tensors 
we obtain the local Casimir forces in all walls by introducing a particular
external configuration. It is shown that this external configuration cannot 
give account of the edge divergences of the local forces. The local form of the 
forces is obtained for three special configurations.

\end{titlepage}
\newpage \baselineskip .37in

\section{Introduction}

In a previous paper, the one-loop renormalization of the anisotropic scalar
model was performed, assuming that the fields were defined in a
d-dimensional Euclidean space where the first $d-1$ coordinates are
unbounded, while the last one lies in the interval $[0,L]$ \cite{fosco}. The
authors analysed the vacuum activity of massive scalar fields assuming
different boundary conditions on the plates, namely Dirichlet-Dirichlet 
$(D-D)$ boundary conditions and also Neumann-Neumann $(N-N)$ boundary
conditions. They obtained two different results. The first one has been
obtained previously by many authors, and is the fact that to renormalize the
theory we have to introduce counterterms as surface interactions. The second
one is the fact that the tadpole graph for $DD$ and for $NN$ have the same 
$z$ dependent part in modulus but with opposite signs. This second result
has been obtained by DeWitt \cite{DeWitt} and also Deutsch and Candelas \cite
{Candelas}. In ref.\cite{fosco}, the authors also investigated the relevance
of this fact to eliminate the surface divergences.

More than thirty years ago, the local version of the Casimir original
problem was performed by Brown and Maclay \cite{brown}. They obtained a
constant stress-energy tensor due to the cancelation of the electric and
magnetic sectors, showing the uniformity of the vacuum of the electromagnetic
field for this configuration. The result is due in part to the particular
field involved and also to the simplicity of the parallel plane geometry. A
similar cancelation can be arranged for the scalar field by computing the
improved stress-energy tensor, but in a more complicated rectangular
geometry with the presence of edges and corners (e.g. within a rectangular
waveguide), we expect an answer strongly non-uniform. The aim of this paper
is to generalize part of the results of Fosco and Svaiter \cite{fosco},
introducing edges in the domain where the fields are defined, calculating
the renormalized stress-energy tensor of a massless scalar field in an
infinitely long rectangular waveguide. As was stressed by Maclay, geometries
with corners present special problems with respect to vacuum energy;
nevertheless these issues have received little attention in the literature 
\cite{macnovo}. It is commonly accepted that, the understanding of the
renormalization of the stress-energy tensor of quantum fields in the presence of
classical boundaries should throw light on the more difficult case, where
there is the added complication of local curvature effects \cite{kay}.

It is well known that there are two quantities which might be expected to
correspond to the total renormalized energy of quantum fields 
\cite{Candelas}. The first one is called the mode sum energy and its definition is 
\begin{equation}
\left<E\right>_{ren}^{mode}=\int_{0}^{\infty }d\omega \,\frac{1}{2}\omega[N(\omega
)-N_{0}(\omega )],  \label{emode}
\end{equation}
where $\frac{1}{2}\omega $ is the zero point energy for each mode, $N(\omega
)d\omega $ is the number of modes with frequencies between $\omega $ and $
\omega +d\omega $ in the presence of boundaries and $N_{0}(\omega )d\omega $
is the corresponding quantity evaluated in empty space. The above equation
gives the renormalized sum of the zero point energy for each mode. The
second one is the volume integral of the renormalized energy density $
\left<E\right>_{ren}^{vol}$ obtained by the Green's function method \cite{brown,green}. A recent investigation of $\left<E\right>_{ren}^{mode}$ 
in rectangular
geometries was given by Svaiter and colaborators \cite{caruso,pro}. A
seminal paper studying this kind of geometry was made by Ambjorn and
Wolfram \cite{ambb}, and more recently Milton and Ng studied the Casimir
effect in $(2+1)$ Maxwell-Chern-Simons electrodynamics in a rectangular
domain \cite{Ng}. Since these definitions deal with integrated quantities,
surface divergence problems do not appear in the calculations. Although
global effects are more accessible to experiments, it is quite important to
understand how the global effect is obtained from the local version. This issue has recently been studied by Actor and Bender \cite{actor,bender}.

In ref. \cite{actor} the author studied the
use of the zeta function method to find the effective action associated with
a scalar field defined in the interior of the infinitely long waveguide,
while in ref. \cite{bender} the authors use the same method to compute the
stress-energy tensor for various rectangular geometries. Using the relation
between the local force density and the discontinuity of the stress-energy
tensor across the boundaries, they computed the local Casimir forces, which
exhibited strong position dependence.

In this paper we are interested in calculating local quantities in the
presence of surfaces and edges. As was stressed by Dowker and Kennedy \cite
{dow} and also Actor and Bender \cite{bender}, to study the local problem in
the infinitely long rectangular waveguide, it is necessary to present the
local form of the analytic continuation of the local zeta function in the
rectangle. Note that our choice of a rectangular cavity is related to the
fact that the modes of the field in this geometric configuration are well
known and an exact calculation can be done.

The organization of the paper is the following: In section II a brief
review of the zeta function method is presented. In section III we calculate
the vacuum expectation value of the canonical and improved stress-energy tensors
associated with a massless scalar field using the zeta function method in the
infinitely long rectangular waveguide. We also show in section III the relation
that exists between the local version and the global version of the Casimir
energy for the waveguide. In section IV, we use the results of
the previous section to compute the local forces. In order to do this we
introduce an external configuration (such that the interior region is the
waveguide) for which the components of the stress-tensors are known 
everywhere. Conclusions are given in section V. In this paper we 
use $\hbar =c=1$.

\section{The canonical and improved stress tensors and the zeta 
         function method}

In this section we will describe the basic procedure to compute the
renormalized vacuum expectation value of the stress-energy tensor for a real
scalar field. Our aproach will be based on the zeta function method.

For a real scalar field defined in a four dimensional spacetime, distorted
by static boundaries, we can use the Fourier standard expansion 
\begin{equation}
\phi (x)=\sum_{n}\frac{1}{\sqrt{2\omega _{n}}}\left[ a_{n}e^{-ix_{0}\omega
_{n}}\phi _{n}(\vec{x})+a_{n}^{\dagger }e^{ix_{0}\omega _{n}}\phi _{n}^{*}
(\vec{x})\right] \ .  \label{fock}
\end{equation}
Assuming that the manifold is static, i.e., that it possesses a timelike Killing
vector field, it is possible to show that there is a complete set of spatial
modes $\left\{ \phi _{n}(\vec{x})\right\} $ satisfying a
Schr\"{o}dinger-like equation 
\begin{equation}
-\Delta \phi _{n}(\vec{x})=\omega _{n}^{2}\phi _{n}(\vec{x}),
\end{equation}
where over these modes we will impose certain boundary conditions. Here we
are concerned only with Dirichlet boundary conditions, although the
generalization to Neumann boundary conditions is straightfoward. Since the
set of modes $\phi _{n}(\vec{x})$ are orthonormal and complete, one then
readily verifies that the equal-time canonical commutation relations imply
the usual commutation relation between annihilation and creation operators of
the quanta of the field.

The main point of interest for us will be the renormalized stress-energy
tensor of the scalar field confined in the interior of the rectangular
infinitely long waveguide. The canonical and improved stress tensors
of a real massless scalar field are given by 
\begin{equation}
T_{\mu \nu }(x)=\frac{1}{2}\left[ \partial _{\mu }\phi \,\partial _{\nu
}\phi +\partial _{\nu }\phi \,\partial _{\mu }\phi -\eta _{\mu \nu }
\partial _{\alpha }\phi \partial ^{\alpha }\phi 
\right]  \label{cst}
\end{equation}
and 
\begin{equation}
\Theta_{\mu \nu }(x)=\frac{1}{3}\left[ \partial _{\mu }\phi \,\partial_{\nu}
\phi +\partial _{\nu }\phi \,\partial _{\mu }\phi 
-\frac1{2}\left(\phi\partial _{\mu } \,\partial_{\nu}\phi
+\partial_{\mu}\partial_{\nu}\phi .\phi + \eta _{\mu \nu }
\partial _{\alpha }\phi \partial ^{\alpha }\phi\right)\right],  
\label{cstIMPR}
\end{equation}
where $\eta _{\mu \nu }$ is the Minkowski metric. One way to write the vacuum 
expectation value of $T_{\mu \nu }(x)$ using eq.(\ref{cst}) is 
\begin{equation}
\left\langle T_{\mu \nu }(x)\right\rangle =\lim_{y\rightarrow x}\frac{1}{2}
\left[ \frac{\partial }{\partial x^{\mu }}\frac{\partial }{\partial y^{\nu }}
+\frac{\partial }{\partial x^{\nu }}\frac{\partial }{\partial y^{\mu }}-\eta
_{\mu \nu }\frac{\partial }{\partial x^{\alpha }}\frac{\partial }{
\partial y_{\alpha }} \right] \left\langle \phi (x)\phi
(y)\right\rangle ,  \label{cst2}
\end{equation}
where $\left\langle \phi (x)\phi (y)\right\rangle $ is the vacuum expectation
value of the product of the fields in two different points. (An equivalent
relation exists for $\left<\Theta_{\mu\nu}\right>$.)
Using the
commutation relations between annihilation and creation operators, the
quantity $\left\langle \phi (x)\phi (y)\right\rangle$ in eq.(\ref{cst2}),
can be written as 
\begin{equation}
\left\langle \phi (x)\phi (y)\right\rangle =\sum_{n}\frac{1}{2\omega _{n}}
\exp (-i\left( x_{0}-y_{0}\right) \omega _{n})\phi _{n}(\vec{x})\phi
_{n}^{*}(\vec{y})\ .  \label{greenf}
\end{equation}
It is clear that $\left\langle T_{\mu \nu }(x)\right\rangle $ can be
obtained from the bilocal sum given by eq.(\ref{greenf}). The bilocal
(spectral) sum in eq.(\ref{greenf}) diverges and needs a regularization and
renormalization procedure. A convenient method is to set $x_{0}=y_{0}$ and
replace $\omega _{n}^{-1}$ in eq.(\ref{greenf}) by $\omega _{n}^{-2s}$ with 
$s$ complex, initially holding for $Re(s)>0$ and sufficiently large to
guarantee convergence even for $\vec{x}=\vec{y}$ , followed by analytic
continuation in $s$.

Let us work with a compact manifold $M$ with or without boundaries. The
diagonal zeta function associated with some elliptic, semi-positive and 
self-adjoint differential operator $D$ will be defined by $\zeta (s|D)$. Let 
$\phi _{n}(x)$ and $\lambda _{n}$ be the spectral decomposition of $D$ in a
complete normal set of eigenfunctions $\phi _{n}(x)$ with eigenvalues
$\lambda _{n}$, i.e. 
\begin{equation}
D\phi _{n}(x)=\lambda _{n}\phi _{n}(x)
\end{equation}
where $\phi _{n}(x)=\left\langle x|n\right\rangle $. Since the eigenfuctions 
$\phi _{n}(x)$ form a complete and normal set it is possible to define the
generalized zeta operator associated with $D$ as 
\begin{equation}
\hat{\zeta}(s|D)=\mu ^{2s}\sum_{n}^{\prime }\frac{\left| n\rangle \langle
n\right| }{\lambda _{n}^{s}},
\end{equation}
where we introduce the parameter $\mu $ with dimensions of mass in order to
have a dimensionless quantity raised to a complex power and the prime sign indicates that the zero eigenvalue of $D$ must be ommited. The generalized
zeta function associated with the operator $D$ is defined by 
\begin{equation}
\zeta (s|D)=\mu ^{2s}\int_{M}d\gamma (x)\left\langle x|D^{-s}|x\right\rangle,
\end{equation}
where $d\gamma (x)$ is the measure on $M$. We have then to consider the bilocal 
zeta function 
\begin{equation}
\zeta \left( s\mid \vec{x},\vec{y}\right) =\mu ^{2s}\sum_{n}\left( \omega
_{n}^{2}\right) ^{-s}\phi _{n}(\vec{x})\phi _{n}^{*}(\vec{y}),  
\label{bizeta}
\end{equation}
which has abscissa of convergence $Re(s)=\frac{3}{2}$. Since the modes 
$\phi _{n}(\vec{x})$ form an orthonormal set then the passage from the local to 
the more familiar global zeta function is straightforward for 
$Re(s)>\frac{3}{2}$. This can be done integrating the bilocal zeta function, 
i.e.
\begin{equation}
\zeta (s)=\mu ^{2s}\int d\gamma (x)\,\zeta\left( s\mid \vec{x},\vec{x}\right)
=\mu ^{2s}\sum_{n}\left( \omega _{n}^{2}\right) ^{-s},\quad Re(s)>\frac{3}{2}.  \label{globalZ}
\end{equation}
A careful analysis of the analytic extension of the global zeta function
associated with some differential operator defined in compact manifold with
or without boundaries can be found in ref. \cite{bn}. Going back to the
local case in the analytic extension of the local zeta function to the whole
complex plane (to the region $Re(s)<\frac{3}{2}$), it will appear poles related
with the geometry of the manifold. For sake of simplicity we will omit the 
$\mu $ factor in the following.

The function given by eq.(\ref{bizeta}) is related to the heat kernel by a
Mellin transform 
\begin{equation}
\zeta \left( s\mid \vec{x},\vec{y}\right) =\frac{1}{\Gamma (s)}
\int_{0}^{\infty }dt\,t^{s-1}K\left( t\mid \vec{x},\vec{y}\right)
\label{z-hk}
\end{equation}
where 
\begin{equation}
K\left( t\mid \vec{x},\vec{y}\right) =\sum_{n}e^{-t\omega _{n}^{2}}\phi _{n}
(\vec{x})\phi _{n}^{*}(\vec{y})  \label{hk}
\end{equation}
is the heat kernel satisfying the same boundary conditions that we choose to
the complete set of modes $\phi _{n}(\vec{x})$. It is possible to express
the vacuum expectation value of the canonical stress tensor given by eq.(\ref
{cst2}) in terms of the modes $\phi _{n}$ and also the frequencies 
$\omega _{n}$. We have
\begin{equation}
\left\langle T_{00}(x)\right\rangle =\frac{1}{4}\sum_{n}\omega _{n}\left|
\phi _{n}\right| ^{2}+\frac{1}{4}\sum_{n}\frac{1}{\omega _{n}}\left| \vec{
\nabla}\phi _{n}\right| ^{2}  \label{T00}
\end{equation}
and 
\begin{equation}
\left\langle T_{ii}(x)\right\rangle =\frac{1}{4}\sum_{n}\omega _{n}\left|
\phi _{n}\right| ^{2}-\frac{1}{4}\sum_{n}\frac{1}{\omega _{n}}\left| 
\vec{\nabla}\phi _{n}\right| ^{2}+\frac{1}{2}
\sum_{n}\frac{1}{\omega _{n}}\left|\partial _{i}\phi _{n}\right| ^{2}  \label{Tii}
\end{equation}
$i$ not summed. It is easy to see that for $\phi _{n}$ real and for plane
waves 
\begin{equation}
\left\langle T_{0i}(x)\right\rangle =0  \label{T0i}
\end{equation}
and finally 
\begin{equation}
\left\langle T_{ij}(x)\right\rangle =\frac{1}{4}\sum_{n}\frac{1}{\omega _{n}}
\left[ \partial _{i}\phi _{n}\partial _{j}\phi _{n}^{*}+\partial _{j}\phi
_{n}\partial _{i}\phi _{n}^{*}\right] \quad i\neq j\ .  \label{Tij}
\end{equation}
For the improved stress tensor we have: 
\begin{equation}
\left\langle \Theta _{00}(x)\right\rangle =\frac{5}{12}\sum_{n}\omega
_{n}\left| \phi _{n}\right| ^{2}+\frac{1}{12}\sum_{n}\frac{1}{\omega _{n}}
\left| \vec{\nabla}\phi _{n}\right| ^{2}
\end{equation}
\begin{equation}
\left\langle \Theta _{ii}(x)\right\rangle =\frac{1}{3}\sum_{n}\frac{1}{
\omega _{n}}\left| \partial _{i}\phi _{n}\right| ^{2}+\frac{1}{12}
\sum_{n}\omega _{n}\left| \phi _{n}\right| ^{2}-\frac{1}{12}\sum_{n}\frac{1}
{\omega _{n}}\left| \vec{\nabla}\phi _{n}\right| ^{2}-\frac{1}{12}\sum_{n}
\frac{1}{\omega _{n}}\left[ \phi _{n}\partial _{i}^{2}\phi _{n}^{*}+\left(
\partial _{i}^{2}\phi _{n}\right) \phi _{n}^{*}\right]
\end{equation}
$i$ not summed,
\begin{equation}
\left\langle \Theta _{0i}(x)\right\rangle =\left\langle
T_{0i}(x)\right\rangle=0 \,,
\end{equation}
\begin{equation}
\left\langle \Theta _{ij}(x)\right\rangle =\frac{1}{6}\sum_{n}\frac{1}{
\omega _{n}}\left[ \partial _{i}\phi _{n}\partial _{j}\phi _{n}^{*}+\partial
_{j}\phi _{n}\partial _{i}\phi _{n}^{*}\right] -\frac{1}{12}\sum_{n}\frac{1}{
\omega _{n}}\left[ \phi _{n}\partial _{i}\partial _{j}\phi _{n}^{*}+\left(
\partial _{i}\partial _{j}\phi _{n}\right) \phi _{n}^{*}\right] \quad i\neq
j\ .
\end{equation}

In the next section we will identify the divergences and the finite parts that appear in the vacuum expectation value of the canonical and the improved stress tensors of a real massless scalar field satisfying Dirichlet boundary 
conditions in all walls of an infinitely long rectangular waveguide.

\section{Canonical and improved stress-energy tensor of a massless scalar
field confined within a rectangular waveguide}

In this section we will apply the local zeta function method to calculate
the renormalized vacuum expectation values of the canonical and improved
stress-energy tensors of a massless scalar field confined within an infinitely 
long rectangular waveguide. Let the waveguide be oriented along the $x_{3}$ 
axis in such a way that the field is defined free in the region 
\begin{equation}
\Omega ={{\bf x}\equiv (x_{1},x_{2},x_{3}):0<x_{1}<a,\quad 0<x_{2}<b}\subset 
{\bf R}^{3},
\end{equation}
with Dirichlet boundary conditions at $x_{1}=0$ and $x_{1}=a$ and also 
$x_{2}=0$ and $x_{2}=b$. The spatial modes are given by: 
\begin{equation}
\phi _{m_{1},m_{2}}\left( \vec{x}\right) =\left( \frac{4}{ab}\right) ^{\frac{
1}{2}}\sin \frac{m_{1}\pi x_{1}}{a}\sin \frac{m_{2}\pi x_{2}}{b}\frac{1}{
\sqrt{2\pi }}e^{ik_{3}x_{3}},  \label{modes}
\end{equation}
with $m_{1,2}=1,2,3,...$ and $-\infty <k_{3}<\infty $. The eigenvalues are
given by 
\begin{equation}
\omega _{n}^{2}=\left( \left( \frac{m_{1}\pi }{a}\right) ^{2}+\left( \frac{
m_{2}\pi }{b}\right) ^{2}+k_{3}^{2}\right) ,
\end{equation}
where $n$ denotes the collective indices $(m_1, m_2, k_3)$. Substituting
eq.(\ref{modes}) in eq.(\ref{hk}) the heat-kernel can be written as: 
\begin{eqnarray}
K\left( t\mid \vec{x},\vec{y}\right) &=&\sum_{m}e^{-t\omega _{m}^{2}}\phi
_{m}\left( \vec{x}\right) \phi _{m}(\vec{y})  \nonumber \\
&=&\frac{1}{2\pi }\left( \frac{4}{ab}\right) \int_{-\infty }^{\infty
}dk_{3}\sum_{m_{1},m_{2}=1}^{\infty }\exp \left\{ -t\left[ \left( \frac{
m_{1}\pi }{a}\right) ^{2}+\left( \frac{m_{2}\pi }{b}\right) ^{2}+\left(
k_{3}\right) ^{2}\right] \right\}  \nonumber \\
\vspace{0.4cm} && \times
\sin (\frac{m_{1}\pi x_{1}}{a})\sin (\frac{m_{2}\pi x_{2}}{b
})\sin (\frac{m_{1}\pi y_{1}}{a})\sin (\frac{m_{2}\pi y_{2}}{b}
)e^{ik_{3}\left( x_{3}-y_{3}\right) }.  \label{hkernel}
\end{eqnarray}
The free spacetime part can be integrated imediately: 
\begin{equation}
\frac{1}{2\pi }\int dk_{3}\,e^{-t\left( k_{3}\right) ^{2}}e^{ik_{3}\left(
x_{3}-y_{3}\right) }=\left( 4\pi t\right) ^{-\frac{1}{2}}\exp \left[ -\frac{
\left( x_{3}-y_{3}\right) ^{2}}{4t}\right] ,
\end{equation}
yielding
\begin{eqnarray}
K\left( t\mid \vec{x},\vec{y}\right) &=&\frac{4}{ab}\left( 4\pi t\right) ^{-
\frac{1}{2}}\exp \left[ -\frac{\left( x_{3}-y_{3}\right) ^{2}}{4t}\right] 
\nonumber \\
&&\times
\sum_{m_{1}=1}^{\infty }\exp \left[ -t\left( \frac{m_{1}\pi }{a}\right)
^{2}\right] \sin (\frac{m_{1}\pi x_{1}}{a})\sin (\frac{m_{1}\pi y_{1}}{a}) 
\nonumber \\
&&\times
\sum_{m_{2}=1}^{\infty }\exp \left[ -t\left( \frac{m_{2}\pi }{b}\right)
^{2}\right] \sin (\frac{m_{2}\pi x_{2}}{b})\sin (\frac{m_{2}\pi y_{2}}{b}).
\end{eqnarray}

Using trigonometric identities and also the Jacobi $\theta $-function
identity 
\begin{equation}
\sum_{m=1}^{\infty }\exp (-m^{2}x)\cos (m2\pi h)=-\frac{1}{2}+\sqrt{\frac{
\pi }{4x}}\sum_{n=-\infty }^{\infty }\exp \left[ -(n+h)^{2}\frac{\pi ^{2}}{x}
\right]
\end{equation}
one finds to the heat-kernel 
\begin{eqnarray}
K\left( t\mid \vec{x},\vec{y}\right) &=&(4\pi t)^{-\frac{3}{2}}\exp \left[ - 
\frac{\left( x_{3}-y_{3}\right) ^{2}}{4t}\right]  \nonumber \\
&& \times
\sum_{n_{1}=-\infty }^{\infty }\left\{ \exp \left[ \frac{-\left[
2n_{1}a+(x_{1}-y_{1})\right] ^{2}}{4t} \right] -\exp \left[ \frac{-\left[
2n_{1}a+(x_{1}+y_{1})\right] ^{2}}{4t} \right] \right\}  \nonumber \\
&& \times
\sum_{n_{2}=-\infty }^{\infty }\left\{ \exp \left[ \frac{-\left[
2n_{2}b+(x_{2}-y_{2})\right] ^{2}}{4t} \right] -\exp \left[ \frac{-\left[
2n_{2}b+(x_{2}+y_{2})\right] ^{2}}{4t} \right] \right\}.  \label{kernel}
\end{eqnarray}

As we discussed before to find the bilocal zeta function we need to perform
the Mellin transform of the heat-kernel given by eq.(\ref{kernel}). All 
terms of eq.(\ref{kernel}) can be integrated using \cite{grad}
\begin{equation}
\int_{0}^{\infty }dt\,t^{s-\frac{5}{2}}\exp (-\frac{A}{t})=A^{s-\frac{3}{2}
}\Gamma (\frac{3}{2}-s).
\end{equation}
After a straightforward calculation we have 
\begin{equation}
\zeta \left( s\mid \vec{x},\vec{y}\right) =\frac{\Gamma (\frac{3}{2}-s)}{
\left( 4\pi \right) ^{\frac{3}{2}}\Gamma (s)}\sum_{n_{1},n_{2}=-\infty
}^{\infty }\left( Z_{1}+Z_{2}+Z_{3}+Z_{4}\right) ,  \label{zeta}
\end{equation}
where $Z_{j}=Z_{j}(n_{1},n_{2},\vec{x},\vec{y}),\quad j=1,2,3,4$, are given by 
\begin{eqnarray}
Z_{1} &=&\left[ \left( n_{1}a+\frac{(x_{1}-y_{1})}{2}\right) ^{2}+\left(
n_{2}b+\frac{(x_{2}-y_{2})}{2}\right) ^{2}+(\frac{x_{3}-y_{3}}{2}
)^{2}\right] ^{s-\frac{3}{2}}  \label{I1} \\
Z_{2} &=&-\left[ \left( n_{1}a+\frac{(x_{1}-y_{1})}{2}\right) ^{2}+\left(
n_{2}b+\frac{(x_{2}+y_{2})}{2}\right) ^{2}+(\frac{x_{3}-y_{3}}{2}
)^{2}\right] ^{s-\frac{3}{2}}  \label{I2a} \\
Z_{3} &=&-\left[ \left( n_{1}a+\frac{(x_{1}+y_{1})}{2}\right) ^{2}+\left(
n_{2}b+\frac{(x_{2}-y_{2})}{2}\right) ^{2}+(\frac{x_{3}-y_{3}}{2}
)^{2}\right] ^{s-\frac{3}{2}}  \label{I3} \\
Z_{4} &=&\left[ \left( n_{1}a+\frac{(x_{1}+y_{1})}{2}\right) ^{2}+\left(
n_{2}b+\frac{(x_{2}+y_{2})}{2}\right) ^{2}+(\frac{x_{3}-y_{3}}{2}%
)^{2}\right] ^{s-\frac{3}{2}}.  \label{I4}
\end{eqnarray}
We see that divergences appear in the local zeta function 
$\zeta(s\mid \vec{x},\vec{y})$ in the limit $\vec{y}\rightarrow \vec{x}.$  
We note that $\zeta (s\mid \vec{x},\vec{x})$ has surface divergences when 
$Re(s)<\frac{3}{2}$ . The term 
$Z_{2}\left( 0,0,\vec{x},\vec{x}\right) =\left(x_{2}\right) ^{2s-3}$, for example, diverges when $x_{2}\rightarrow 0$ in this case.

In order to calculate the components of 
$\left\langle T_{\mu \nu }(\vec{x})\right\rangle $ we have to evaluate 
the mode sums given by eqs.(\ref{T00})-(\ref{Tij}). One then readily 
verifies that
\begin{equation}
\sum_{n}\omega _{n}\left| \phi _{n}\right| ^{2}=
\zeta(s=-\frac{1}{2}\mid \vec{x},
\vec{x})=-\frac{1}{16\pi ^{2}}F_{0}(\vec{x}),  \label{w}
\end{equation}
where the expression for $F_{0}(\vec{x})$ is given by 
\begin{eqnarray}
 F_{0}(\vec{x}) &=&\sum_{n_{1},n_{2}=-\infty }^{\infty }\left[
 (n_{1}a)^{2}+(n_{2}b)^{2}\right] ^{-2}+  \nonumber \\
&&\hspace{1cm}
 -\left[ (n_{1}a)^{2}+\left( n_{2}b+x_{2}\right) ^{2}\right] ^{-2}+ 
\nonumber \\
&&\hspace{1cm}
-\left[ \left( n_{1}a+x_{1}\right) ^{2}+\left( n_{2}b\right) ^{2}\right]
^{-2}+  \nonumber \\
&&\hspace{1cm}
+\left[ \left( n_{1}a+x_{1}\right) ^{2}+\left( n_{2}b+x_{2}\right)
^{2}\right] ^{-2}.  \label{F0}
\end{eqnarray}

The other terms that we need are given by: 
\begin{equation}
\sum_{n}\frac{1}{\omega _{n}}|\partial _{i}\phi _{n}|^{2}=\lim_{\vec{y}
\rightarrow \vec{x}}\frac{\partial }{\partial x^{i}}\frac{\partial }{
\partial y^{i}}\zeta(s=\frac{1}{2}\mid \vec{x},\vec{y}).  \label{Zi}
\end{equation}
Substituting eq.(\ref{zeta}) in eq.(\ref{Zi}) for $i=1,$ we have:
\begin{equation}
\sum_{n}\frac{1}{\omega _{n}}\left| \partial _{1}\phi _{n}\right| ^{2}=-
\frac{1}{4\pi ^{2}}D_{1}(\vec{x})+\frac{1}{16\pi ^{2}}F_{1}(\vec{x}),
\label{d1}
\end{equation}
where the functions $D_{1}(\vec{x})$ and F$_{1}\left( \vec{x}\right) $ are
defined by 
\begin{eqnarray}
D_{1}(\vec{x}) &=&\sum_{n_{1},n_{2}=-\infty }^{\infty }\left[
(n_{1}a)^{2}+(n_{2}b)^{2}\right] ^{-3}\left[ n_{1}a\right] ^{2}  \nonumber \\
&&\hspace{1cm}
-\left[ (n_{1}a)^{2}+\left( n_{2}b+x_{2}\right) ^{2}\right] ^{-3}\left[
n_{1}a\right] ^{2}+  \nonumber \\
&&\hspace{1cm}
+\left[ \left( n_{1}a+x_{1}\right) ^{2}+\left( n_{2}b\right) ^{2}\right]
^{-3}\left[ n_{1}a+x_{1}\right] ^{2}+  \nonumber \\
&&\hspace{1cm}
-\left[ \left( n_{1}a+x_{1}\right) ^{2}+\left( n_{2}b+x_{2}\right)
^{2}\right] ^{-3}\left[ n_{1}a+x_{1}\right] ^{2}  \label{D1}
\end{eqnarray}
and
\begin{eqnarray}
F_{1}(\vec{x}) &=&\sum_{n_{1},n_{2}=-\infty }^{\infty }\left[
(n_{1}a)^{2}+(n_{2}b)^{2}\right] ^{-2}+  \nonumber \\
&&\hspace{1cm}
-\left[ (n_{1}a)^{2}+\left( n_{2}b+x_{2}\right) ^{2}\right] ^{-2}+ 
\nonumber \\
&&\hspace{1cm}
+\left[ \left( n_{1}a+x_{1}\right) ^{2}+\left( n_{2}b\right) ^{2}\right]
^{-2}+  \nonumber \\
&&\hspace{1cm}
-\left[ \left( n_{1}a+x_{1}\right) ^{2}+\left( n_{2}b+x_{2}\right)
^{2}\right] ^{-2}.  \label{F1}
\end{eqnarray}
For $i=2$, 
\begin{equation}
\sum_{n}\frac{1}{\omega _{n}}\left| \partial _{2}\phi _{n}\right| ^{2}=-
\frac{1}{4\pi ^{2}}D_{2}(\vec{x})+\frac{1}{16\pi ^{2}}F_{2}(\vec{x}),
\label{d2}
\end{equation}
where the functions $D_{2}(\vec{x})$ and $F_{2}(\vec{x})$ are defined by 
\begin{eqnarray}
D_{2}(\vec{x}) &=&\sum_{n_{1},n_{2}=-\infty }^{\infty }\left[
(n_{1}a)^{2}+(n_{2}b)^{2}\right] ^{-3}\left[ n_{2}b\right] ^{2}  \nonumber \\
&&\hspace{1cm}
+\left[ (n_{1}a)^{2}+\left( n_{2}b+x_{2}\right) ^{2}\right] ^{-3}\left[
n_{2}b+x_{2}\right] ^{2}+  \nonumber \\
&&\hspace{1cm}
-\left[ \left( n_{1}a+x_{1}\right) ^{2}+\left( n_{2}b\right) ^{2}\right]
^{-3}\left[ n_{2}b\right] ^{2}+  \nonumber \\
&&\hspace{1cm}
-\left[ \left( n_{1}a+x_{1}\right) ^{2}+\left( n_{2}b+x_{2}\right)
^{2}\right] ^{-3}\left[ n_{2}b+x_{2}\right] ^{2}  \label{D2}
\end{eqnarray}
and
\begin{eqnarray}
F_{2}(\vec{x}) &=&\sum_{n_{1},n_{2}=-\infty }^{\infty }\left[
(n_{1}a)^{2}+(n_{2}b)^{2}\right] ^{-2}+  \nonumber \\
&&\hspace{1cm}
+\left[ (n_{1}a)^{2}+\left( n_{2}b+x_{2}\right) ^{2}\right] ^{-2}+ 
\nonumber \\
&&\hspace{1cm}
-\left[ \left( n_{1}a+x_{1}\right) ^{2}+\left( n_{2}b\right) ^{2}\right]
^{-2}+  \nonumber \\
&&\hspace{1cm}
-\left[ \left( n_{1}a+x_{1}\right) ^{2}+\left( n_{2}b+x_{2}\right)
^{2}\right] ^{-2}.  \label{F2}
\end{eqnarray}
For $i=3$ 
\begin{equation}
\sum_{n}\frac{1}{\omega _{n}}\left| \partial _{3}\phi _{n}\right| ^{2}=\frac{1}{16\pi ^{2}}F_{0}(\vec{x}).  \label{d3}
\end{equation}
We still need to calculate 
\begin{equation}
\sum_{n}\frac{1}{\omega _{n}}\partial _{i}\phi _{n}\partial _{j}\phi
_{n}^{*}=\lim_{\vec{y}\rightarrow \vec{x}}\frac{\partial }{\partial x^{i}}
\frac{\partial }{\partial y^{i}}\zeta(s=\frac{1}{2}\mid \vec{x},\vec{y}).
\end{equation}
For i=2 and j=1, we have 
\begin{equation}
\sum_{n}\frac{1}{\omega _{n}}\partial _{2}\phi _{n}\partial _{1}\phi
_{n}^{*}=\sum_{n}\frac{1}{\omega _{n}}\partial _{1}\phi _{n}\partial _{2}\phi
_{n}^{*}=
-\frac{F_{21}(\vec{x})}{4\pi ^{2}}\,,  \label{d21}
\end{equation}
where the function $F_{21}(\vec{x})$ is defined by 
\begin{eqnarray}
F_{21}(\vec{x}) &=&\sum_{n_{1},n_{2}=-\infty }^{\infty } 
\left[ (n_{1}a)^{2}+(n_{2}b)^{2}\right] ^{-3}\left[ n_{2}b\right] \left[
n_{1}a\right] +  \nonumber \\
&&\hspace{1cm}
-\left[ \left( n_{2}b+x_{2}\right) ^{2}+\left( n_{1}a\right) ^{2}\right]
^{-3}\left[ n_{2}b+x_{2}\right] \left[ n_{1}a\right] +  \nonumber \\
&&\hspace{1cm}
+\left[ \left( n_{1}a+x_{1}\right) ^{2}+\left( n_{2}b\right) ^{2}\right]
^{-3}\left[ n_{1}a+x_{1}\right] \left[ n_{2}b\right] +  \nonumber \\
&&\hspace{1cm}
-\left[ \left( n_{1}a+x_{1}\right) ^{2}+\left( n_{2}b+x_{2}\right)
^{2}\right] ^{-3}\left[ n_{1}a+x_{1}\right] \left[ n_{2}b+x_{2}\right] 
\nonumber \\
&=&
-\sum_{n_{1},n_{2}=-\infty }^{\infty }\left[ \left( n_{1}a+x_{1}\right)
^{2}+\left( n_{2}b+x_{2}\right) ^{2}\right] ^{-3}\left[ n_{1}a+x_{1}\right]
\left[ n_{2}b+x_{2}\right],  \label{F21}
\end{eqnarray}
because the first three summands are odd in one index.

For i=3 and j=1 and for i=3 and j=2, we have 
\begin{equation}
\sum_{n}\frac{1}{\omega _{n}}\partial _{3}\phi _{n}\partial _{1}\phi
_{n}^{*}=\sum_{n}\frac{1}{\omega _{n}}\partial _{3}\phi _{n}\partial
_{2}\phi _{n}^{*}=0.  \label{d31}
\end{equation}
To obtain the components of the improved stress tensor we need to calculate 
\begin{equation}
\sum_{n}\frac{1}{\omega _{n}}\phi _{n}\partial _{i}^{2}\phi _{n}^{*}=\sum_{n}
\frac{1}{\omega _{n}}\left( \partial _{i}^{2}\phi _{n}\right) \phi
_{n}^{*}=\lim_{\vec{y}\rightarrow \vec{x}}\left( \frac{\partial }{\partial
y^{i}}\right) ^{2}\zeta(s=\frac{1}{2}\mid \vec{x},\vec{y}).
\end{equation}
For $i=1$
\begin{equation}
\sum_{n}\frac{1}{\omega _{n}}\phi _{n}\partial _{1}^{2}\phi _{n}^{*}=
\frac{1}{4\pi ^{2}}D_{11}(\vec{x})-\frac{1}{16\pi ^{2}}F_{0}(\vec{x})\,,
\end{equation}
where the function $D_{11}(\vec{x})$ is defined by 
\begin{eqnarray}
D_{11}(\vec{x}) &=&\sum_{n_{1},n_{2}=-\infty }^{\infty }\{\left[
(n_{1}a)^{2}+(n_{2}b)^{2}\right] ^{-3}\left[ n_{1}a\right] ^{2}  \nonumber \\
&&\hspace{1cm}
-\left[ (n_{1}a)^{2}+\left( n_{2}b+x_{2}\right) ^{2}\right] ^{-3}\left[
n_{1}a\right] ^{2}+  \nonumber \\
&&\hspace{1cm}
-\left[ \left( n_{1}a+x_{1}\right) ^{2}+\left( n_{2}b\right) ^{2}\right]
^{-3}\left[ n_{1}a+x_{1}\right] ^{2}+  \nonumber \\
&&\hspace{1cm}
+\left[ \left( n_{1}a+x_{1}\right) ^{2}+\left( n_{2}b+x_{2}\right)
^{2}\right] ^{-3}\left[ n_{1}a+x_{1}\right] ^{2}\}.  \label{D11}
\end{eqnarray}
For $i=2$
\begin{equation}
\sum_{n}\frac{1}{\omega _{n}}\phi _{n}\partial _{2}^{2}\phi _{n}^{*}=\frac{1
}{4\pi ^{2}}D_{22}(\vec{x})-\frac{1}{16\pi ^{2}}F_{0}(\vec{x}),
\end{equation}
where the function $D_{22}(\vec{x})$ can be obtained from $D_{11}(\vec{x})$
with the change $x_{1}\leftrightarrow x_{2}$ and 
$n_{1}a\leftrightarrow n_{2}b$.
For $i=3$
\begin{equation}
\sum_{n}\frac{1}{\omega _{n}}\phi _{n}\partial _{3}^{2}\phi _{n}^{*}=
-\frac{1}{16\pi ^{2}}F_{0}(\vec{x}).
\end{equation}

Finally 
\begin{equation}
\sum_{n}\frac{1}{\omega _{n}}\phi _{n}\partial _{i}\partial _{j}\phi
_{n}^{*}=\lim_{\vec{y}\rightarrow \vec{x}}\frac{\partial }{\partial y^{i}}
\frac{\partial }{\partial y^{j}}\zeta(s=\frac{1}{2}\mid \vec{x},\vec{y}).
\end{equation}
For $i=1$ and $j=2$
\begin{equation}
\sum_{n}\frac{1}{\omega _{n}}\phi _{n}\partial _{1}\partial _{2}\phi
_{n}^{*}=\sum_{n}\frac{1}{\omega _{n}}\phi _{n}\partial _{2}\partial _{1}\phi
_{n}^{*}=-\frac{F_{12}(\vec{x})}{4\pi ^{2}}.
\end{equation}
Substituting the results of eqs.(\ref{w}), (\ref{d1}), (\ref{d2}), (\ref{d3}), (\ref{d21}) and (\ref{d31}) in eqs. (\ref{T00}-\ref{Tij}), we obtain:
\begin{eqnarray}
\left\langle T_{00}(\vec{x})\right\rangle &=&-\frac{1}{16\pi ^{2}}\left( D_{1}
(\vec{x})+D_{2}(\vec{x})\right) +\frac{1}{64\pi ^{2}}\left( F_{1}(\vec{x}
)+F_{2}(\vec{x})\right)  \nonumber \\
&=&\left\langle T_{00}(\vec{x})\right\rangle _{B}+\left\langle T_{00}(\vec{x}
)\right\rangle _{F},
\label{B+F}
\end{eqnarray}
where B and F mean the boundary divergent and finite part respectively. In
explicit form: 
\begin{eqnarray}
\left\langle T_{00}(\vec{x})\right\rangle _{B} &=&\frac{1}{32\pi ^{2}}
\{\left[ \left( x_{1}\right) ^{2}+\left( x_{2}\right) ^{2}\right]
^{-2}+\left[ \left( a-x_{1}\right) ^{2}+\left( b-x_{2}\right) ^{2}\right]
^{-2}+  \nonumber \\
&&\hspace{1cm}
+\left[ \left( x_{1}\right) ^{2}+\left( b-x_{2}\right) ^{2}\right]
^{-2}+\left[ \left( a-x_{1}\right) ^{2}+\left( x_{2}\right) ^{2}\right]
^{-2}\}+  \nonumber \\
&&-\frac{1}{16\pi ^{2}}\{\left[ x_{2}\right] ^{-4}+\left[ b-x_{2}\right]
^{-4}+\left[ x_{1}\right] ^{-4}+\left[ a-x_{1}\right] ^{-4}\}  \label{T00B},
\label{t00d}
\end{eqnarray}
\begin{eqnarray}
\left\langle T_{00}(\vec{x})\right\rangle _{F} &=&-\frac{1}{32\pi ^{2}}
\left(\right.
\sum_{\left( n_{1},n_{2}\right) \neq \left( 0,0\right) }\left[
(n_{1}a)^{2}+(n_{2}b)^{2}\right] ^{-2}+  \nonumber \\
&&-\sum_{\left( n_{1},n_{2}\right) \neq \left( 0,0\right) ,\left(
-1,-1\right) ,(0,-1),(-1,0)}\left[ \left( n_{1}a+x_{1}\right) ^{2}+\left(
n_{2}b+x_{2}\right) ^{2}\right] ^{-2}  +  \nonumber \\
&&+2\sum_{\left( n_{1},n_{2}\right) \neq \left( 0,0\right) ,\left(
0,-1\right) }\left[ (n_{1}a)^{2}+\left( n_{2}b+x_{2}\right) ^{2}\right]
^{-3}([n_{2}b+x_{2}]^{2}-\left[ n_{1}a\right] ^{2})+  \nonumber \\
&&+2\sum_{\left( n_{1},n_{2}\right) \neq \left( 0,0\right) ,\left(
-1,0\right) }\left[ \left( n_{1}a+x_{1}\right) ^{2}+\left( n_{2}b\right)
^{2}\right] ^{-3}(\left[ n_{1}a+x_{1}\right] ^{2}-\left[ n_{2}b\right]
^{2})\left.\right).  
\label{T00F}
\end{eqnarray}
The restriction in the first sum above accounts for the exclusion of the 
free space divergent term.
We see that $\left\langle T_{00}(\vec{x})\right\rangle _{B}$ diverges in all
walls, i.e., $x_{1}=0,a$ and $x_{2}=0,b$ and in all edges, $\left(
x_{1},x_{2}\right) =\left( 0,0\right) $,$\left( a,0\right) $,$\left(
0,b\right) $,$\left( a,b\right) $ of the waveguide. The other components are:
\begin{equation}
\left\langle T_{11}(\vec{x})\right\rangle =-\frac{1}{32\pi ^{2}}F_{0}(\vec{x}
)-\frac{1}{16\pi ^{2}}\left( D_{1}(\vec{x})-D_{2}(\vec{x})\right) +\frac{1}{
64\pi ^{2}}\left( F_{1}(\vec{x})-F_{2}(\vec{x})\right)
\end{equation}
\begin{equation}
\left\langle T_{22}(\vec{x})\right\rangle =-\frac{1}{32\pi ^{2}}F_{0}(\vec{x}
)+\frac{1}{16\pi ^{2}}\left( D_{1}(\vec{x})-D_{2}(\vec{x})\right) -\frac{1}{
64\pi ^{2}}\left( F_{1}(\vec{x})-F_{2}(\vec{x})\right)
\label{t22}
\end{equation}
\begin{equation}
\left\langle T_{33}(\vec{x})\right\rangle =-\left\langle T_{00}(\vec{x}
)\right\rangle
\end{equation}
\begin{equation}
\left\langle T_{12}(\vec{x})\right\rangle =\left\langle T_{21}(\vec{x}
)\right\rangle =-\frac{1}{8\pi ^{2}}F_{12}(\vec{x})
\end{equation}
\begin{equation}
\left\langle T_{23}(\vec{x})\right\rangle =\left\langle T_{32}(\vec{x}
)\right\rangle =\left\langle T_{31}(\vec{x})\right\rangle =\left\langle
T_{13}(\vec{x})\right\rangle =\left\langle T_{0i}(\vec{x})\right\rangle
=\left\langle T_{i0}(\vec{x})\right\rangle =0.
\end{equation}
It can be verified by writing
these quantities explicitly that all the non-zero components have boundary
divergences in all walls and all edges also. As has been remarked previously
by many authors the divergences that appear in some components of the vacuum
expectation value of the stress tensor are related with the unphysical
boundary conditions imposed on the field. We can understand why the
renormalized stress tensor becomes infinite on the boundary. This is related
with the uncertainty relation between the field and the canonical conjugate
momentum associated with the field \cite{or,Zy,ls}.

For the improved stress tensor we have 
\begin{eqnarray}
\left\langle \Theta _{00}(\vec{x})\right\rangle &=&-\frac{1}{48\pi ^{2}}F_{0}
(\vec{x})-\frac{1}{48\pi ^{2}}\left( D_{1}(\vec{x})+D_{2}(\vec{x})\right) +
\frac{1}{192\pi ^{2}}\left( F_{1}(\vec{x})+F_{2}(\vec{x})\right)  \nonumber \\
&=&\frac1{3}\left<T_{00}(\vec{x})\right>-\frac1{48 \pi^2}F_0(\vec{x})
\nonumber \\
&=&\left\langle \Theta _{00}(\vec{x})\right\rangle _{B}+\left\langle \Theta
_{00}(\vec{x})\right\rangle _{F}.
\end{eqnarray}
Explicitly: 
\begin{eqnarray}
\left\langle \Theta _{00}(\vec{x})\right\rangle _{B} &=&-\frac{1}{96\pi ^{2}}
\{\left[ \left( x_{1}\right) ^{2}+\left( x_{2}\right) ^{2}\right]
^{-2}+\left[ \left( a-x_{1}\right) ^{2}+\left( b-x_{2}\right) ^{2}\right]
^{-2}+  \nonumber \\
&&\hspace{1cm}
+\left[ \left( x_{1}\right) ^{2}+\left( b-x_{2}\right) ^{2}\right]
^{-2}+\left[ \left( a-x_{1}\right) ^{2}+\left( x_{2}\right) ^{2}\right]
^{-2}\}
\end{eqnarray}
\begin{eqnarray}
\left\langle \Theta _{00}(\vec{x})\right\rangle _{F} &=&-\frac{1}{32\pi ^{2}}
\sum\left[
(n_{1}a)^{2}+(n_{2}b)^{2}\right] ^{-2}-
\frac1{96\pi^2}\sum
\left[(n_1a + x_1)^2+(n_2b + x_2)^2 \right]^{-2} \nonumber \\
&+&\frac1{24\pi^2}\left(\sum
(n_2b)^2\left[(n_1a + x_1)^2+(n_2b)^2 \right]^{-3} + 
\sum(n_1a)^2\left[(n_1a)^2+(n_2b + x_2)^2 \right]^{-3}\right). \nonumber \\
\label{TETA00F}
\end{eqnarray}
We have to exclude: in the first sum the term $(n_1,n_2)= (0,0)$,
in the second sum the terms $(n_1,n_2)= (0,0),(-1,-1),(-1,0),(0,-1)$.

We see that $\left\langle \Theta _{00}(\vec{x})\right\rangle $ has no wall
divergences but only edge ones, as pointed in \cite{actor,bender}. 
Although we cannot associate a curvature length to the edges, they seem to have a similar behaviour, since the divergences associated with them still remain even for the conformally coupled scalar field. The other components are:
\begin{equation}
\left\langle \Theta _{11}(\vec{x})\right\rangle =-\frac{1}{16\pi ^{2}}D_{1}(
\vec{x})+\frac{1}{48\pi ^{2}}D_{2}(\vec{x})-\frac{1}{24\pi ^{2}}D_{11}(\vec{x}
)+\frac{1}{64\pi ^{2}}F_{1}(\vec{x})-\frac{1}{192\pi ^{2}}F_{2}(\vec{x})
\end{equation}
\begin{equation}
\left\langle \Theta _{22}(\vec{x})\right\rangle =-\frac{1}{16\pi ^{2}}D_{2}(
\vec{x})+\frac{1}{48\pi ^{2}}D_{1}(\vec{x})-\frac{1}{24\pi ^{2}}D_{22}(\vec{x}
)+\frac{1}{64\pi ^{2}}F_{2}(\vec{x})-\frac{1}{192\pi ^{2}}F_{1}(\vec{x})
\end{equation}
\begin{equation}
\left\langle \Theta _{33}(\vec{x})\right\rangle =-\left\langle \Theta _{00}
(\vec{x})\right\rangle
\end{equation}
\begin{equation}
\left\langle \Theta _{12}(\vec{x})\right\rangle =\left\langle \Theta _{21}(
\vec{x})\right\rangle =-\frac{1}{24\pi ^{2}} F_{12}(\vec{x})
\end{equation}
\begin{equation}
\left\langle \Theta _{23}(\vec{x})\right\rangle =\left\langle \Theta _{32}(
\vec{x})\right\rangle =\left\langle \Theta _{31}(\vec{x})\right\rangle
=\left\langle \Theta _{13}(\vec{x})\right\rangle =\left\langle \Theta _{0i}(
\vec{x})\right\rangle =\left\langle \Theta _{i0}(\vec{x})\right\rangle =0.
\end{equation}
Again, by writing these quantities explicitly, it is easy to see that all non-zero components of the improved stress tensor are free of wall divergences 
but have edge divergences.

We are now interested in comparing the local calculation of the energy density
$\left<T_{00}(\vec{x})\right>$ with the more familiar global one. In this way, 
let us now calculate the global energy inside the waveguide by integrating
the energy density $\left< T_{00}(\vec{x})\right>$ given by eq.(\ref{B+F}) 
over the cavity, $0\leq x_1 \leq a, 0 \leq x_2 \leq b$. Despite the fact that
a closed form of the double sums in $\left<T_{00}(\vec{x})\right>$ are 
presently not known, it is possible to calculate its integrals over the spatial 
region of the waveguide. We shall devide this total energy by the area of the 
cross-section of the waveguide $a\times b$, and it is usually refered to also 
as energy density, although this comes from an integrated quantity per unit 
area and is not actually a density in the sense of a local quantity, this one 
legitimately represented by $\left< T_{00}(\vec{x})\right>$.  We shall assume, 
for the global computation, that the field exists only inside the cavity.

Clearly the following expression:
\begin{equation}
 \int_0^a dx_1 \int_0^b dx_2 \left< T_{00}(\vec{x})\right> = 
 \int\int_{cavity} \left< T_{00}(\vec{x})\right> =
 \int\int_{cavity} \left< T_{00}(\vec{x})\right>_B +
 \int\int_{cavity} \left< T_{00}(\vec{x})\right>_F
\label{cavity}
\end{equation}
diverges because of the first term: $\left< T_{00}(\vec{x})\right>_B$ is divergent on the walls and edges. So let us treat the second term, in which 
$\left< T_{00}(\vec{x})\right>_F$ is given by eq.(\ref{T00F}) and is finite.  The integral of the first term of eq.(\ref{T00F}) gives:
\begin{equation}
 -\frac{ab}{32\pi^2}Z(2|a,b),
\end{equation}
where, in the notation of \cite{actor}
\begin{equation}
 Z(2|a,b)=\sum_{(n_1,n_2)\neq (0,0)}\left[(n_1a)^2 + (n_2b)^2\right]^{-2}.
\end{equation}
Again following the notation of \cite{actor}, the integral of the second 
term of eq.(\ref{T00F}) is given by:
\begin{eqnarray}
 \frac1{32\pi^2}\int_0^a dx_1 \int_0^b dx_2 \,\,\zeta_F(2|x_1,x_2)&=&
 \frac1{32\pi^2}\int_{ext} \zeta_B(2|x_1,x_2)=\nonumber \\
 &=&\frac1{8\pi^2}\left[\int_a^{\infty}\int_b^{\infty}  
                      + \int_a^{\infty}\int_0^b
                      + \int_0^a\int_b^{\infty}\right]
 \,dx dy \frac1{(x^2 + y^2)^2},
\label{zetaExt}
\end{eqnarray}
where
\begin{equation}
 \zeta_F(2|x_1,x_2)=\sum_{(n_1,n_2)\neq (0,0)(0,-1)(-1,0)(-1,-1)}
 \left[(n_1a + x_1)^2 + (n_2b + x_2)^2\right]^{-2}
\end{equation}
and the right-hand side is each one of the divergent edge terms integrated 
over the appropriate quadrant {\it outside} the cavity, i.e., away of the 
points where they diverge, and thus eq.(\ref{zetaExt}) is also a finite 
contribution (see \cite{actor} for further explanations).

The integration of the third (and fourth) term of eq.(\ref{T00F}) is not difficult:
\begin{eqnarray}
 &&-\frac1{16\pi^2} \int_0^a dx_1 \int_0^b dx_2 \sum_{(n_1,n_2)\neq (0,0)(0,-1)}\left[(n_1a)^2 + (n_2b + x_2)^2\right]^{-3}
 \left[(n_2b + x_2)^2-(n_1a)^2\right] = \nonumber \\
 && = -\frac1{16\pi^2} \int_0^a dx_1 \int_0^b dx_2 \left(\sum_{n_2\neq 0,-1}\frac1{(n_2b + x_2)^4} + \right.\nonumber \\
 && \left.  \hspace{4cm} + \sum_{n_2=-\infty,n_1\neq 0}^{\infty}\left[(n_1a)^2 
 + (n_2b + x_2)^2\right]^{-3}
 \left[(n_2b + x_2)^2-(n_1a)^2\right] \right).
\label{3term}
\end{eqnarray}
The first integral above was also calculated in \cite{actor}:
\begin{equation}
 -\frac1{16\pi^2} \int_0^a dx_1 \int_0^b dx_2 \,\,
 \sum_{n_2\neq 0,-1}\frac1{(n_2b + x_2)^4} = 
 -\frac{a}{16\pi^2} \int_{ext} \zeta_B(2|0,x_2),
\label{3termExt}
\end{equation}
where
\begin{equation}
 \int_{ext} \zeta_B(2|0,x_2) = \int_b^{\infty} dx_2 \,\,\frac1{x_2^4} + \int_{-\infty}^0 dx_2 \,\,\frac1{(b - x_2)^4}
\label{3termExt1}
\end{equation}
is also finite. The other term gives
\begin{eqnarray}
 -\frac1{16\pi^2} \int_0^a dx_1 \int_0^b dx_2 
 \sum_{n_2=-\infty,n_1\neq 0}^{\infty}\left[(n_1a)^2 + (n_2b + x_2)^2\right]^{-3} \left[(n_2b + x_2)^2-(n_1a)^2\right] 
 &=& \nonumber \\
 && \hspace{-8cm} =  
 -\frac{a}{4\pi^2} \sum_{n=1}^{\infty} \int_0^{\infty} dy \,\,
 \frac{y^2 - (na)^2}{\left[y^2 + (na)^2\right]^{3}} 
 \nonumber \\
 && \hspace{-8cm} = + \frac1{32\pi a^2} \,\,\zeta(3),
\label{3term2}
\end{eqnarray}
where 
$$
 \zeta(s)=\sum_{n=1}^{\infty} n^{-s}
$$ 
is the usual Riemann zeta function, and use has been made of
the integral \cite{grad} 
$$ 
 \int_0^{\infty} dx \,\,\frac{x^{\mu-1}}{\left[1 + \beta x\right]^{\nu}} =   
 \beta^{-\mu}\,\,\frac{\Gamma(\mu)\Gamma(\nu-\mu)}{\Gamma(\nu)},
 \hspace{2cm} \left| \arg\beta\right| < \pi;\,\, \Re \nu>\Re \mu > 0. 
$$

Gathering all previous results we have that:
\begin{equation}
 \int\int_{cavity} \left<T_{00}(\vec{x})\right>_F = -\frac{ab}{32\pi^2}Z(2|a,b)
 + \frac1{32\pi}\zeta(3)\left(\frac1{a^2} + \frac1{b^2}\right) 
 + \int_{ext} \left<T_{00}(\vec{x})\right>_B,
\label{EFinite}
\end{equation}
where
\begin{equation}
 \int_{ext} \left<T_{00}(\vec{x})\right>_B
 = -\frac1{16\pi^2}\left(a\int_{ext}\zeta_B(2|0,x_2) +    
 b\int_{ext}\zeta_B(2|x_1,0)\right)
 +\frac1{32\pi^2}\int_{ext}\zeta_B(2|x_1,x_2)
\end{equation}
is finite, because it is the sum of each of the wall and edge divergent terms 
integrated {\it outside} the cavity, i.e., far from the spatial points where 
they diverge. Eq.(\ref{EFinite}) can be written as:
\begin{equation}
 \frac1{ab}\int\int_{cavity} \left<T_{00}(\vec{x})\right>_F = 
 E_C(a,b) + \frac1{ab} \int_{ext} \left<T_{00}(\vec{x})\right>_B,
\label{LocGlob}
\end{equation}
where $E_C(a,b)$ is the global Casimir energy divided by
the cross-section area $a\times b$ for the waveguide, in agreement with
\cite{actor} (in fact, Actor's definition of $V_{eff}$ is twice the
usual one). We can add the same infinite term 
$$
 \int\int_{cavity} \left<T_{00}(\vec{x})\right>_B
$$
to both sides of the equation above, obtaining:
\begin{equation}
 \frac1{ab}\int\int_{cavity} \left<T_{00}(\vec{x})\right> \, = \,\,
 E_C(a,b)\, + \,\frac1{ab} \int_{all\,\,space} \left<T_{00}(\vec{x})\right>_B,
\end{equation}
where the last integral above is an infinite constant independent of the
cavity dimensions $a,b$.  In global calculations one usually discards
this infinite constant because it does not give rise to forces.
Discarding this infinite constant, one obtains from the expression above
the total Casimir energy per unit area inside the waveguide with Dirichlet 
boundary conditions in all walls. It can be shown that the improved 
stress-tensor yields the same Casimir energy per unit area $E_C(a,b)$:
\begin{equation}
 \frac1{ab}\int\int_{cavity} \left<\Theta_{00}(\vec{x})\right> \, = \,\,
 E_C(a,b)\, + \,\frac1{ab} 
 \int_{all\,\,space} \left<\Theta_{00}(\vec{x})\right>_B.
\end{equation}

It is known that the sign of the global Casimir
energy is dependent on the relative size of $a$ and $b$. For example, for
the square waveguide $a=b$ a positive value for $E_C(a,b)$ is found. Because
this is a symmetric configuration, an equal total outward force appears acting
on each of the four walls, which tends to make the cavity expand. 

\begin{figure}
\centerline {\epsfig{figure=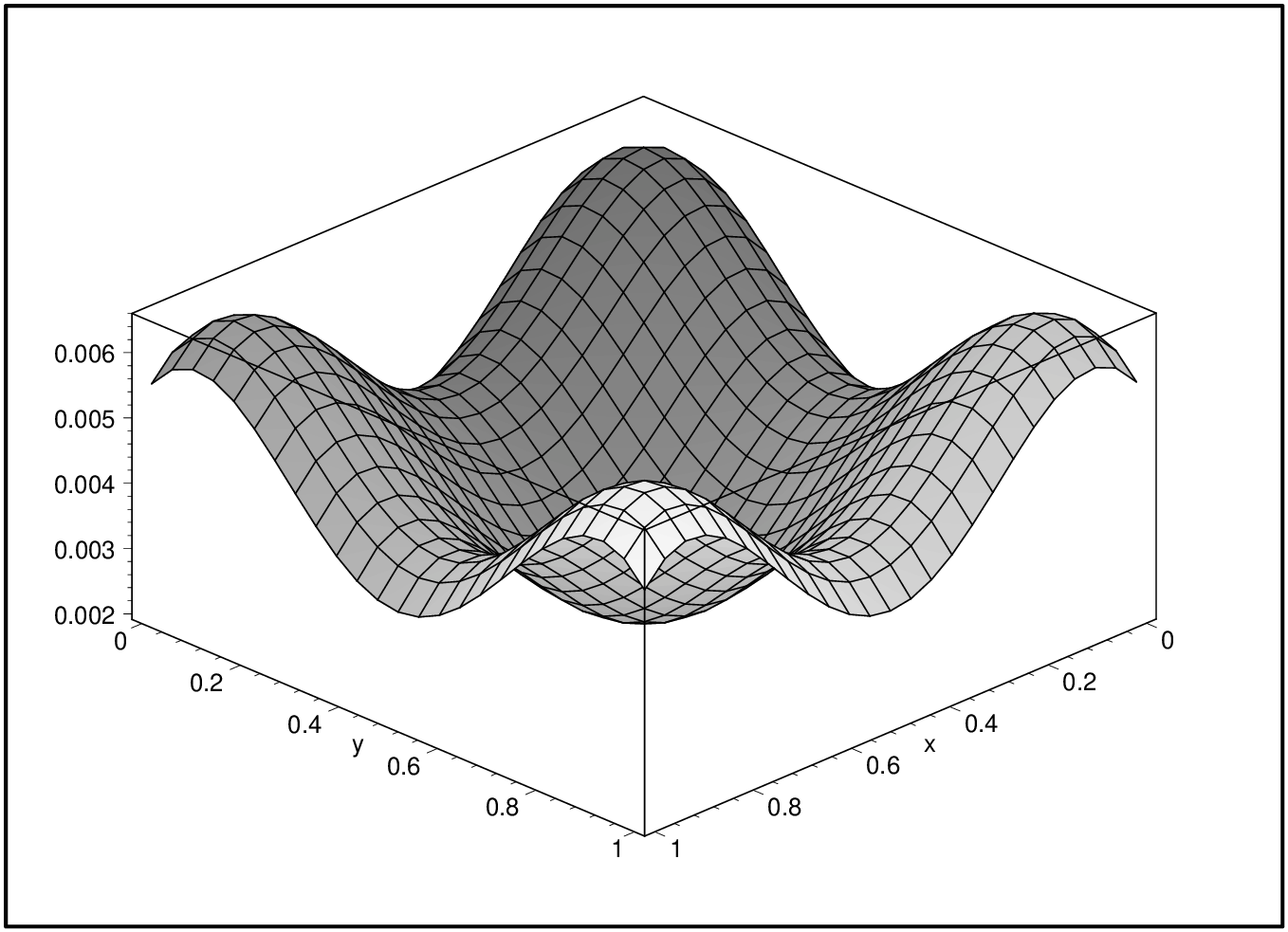, angle=270, width=11cm} }
\caption{Renormalized local energy density of the minimally coupled scalar field in the interior of the square waveguide.}
\label{T00S}
\end{figure}

\begin{figure}
\centerline {\epsfig{figure=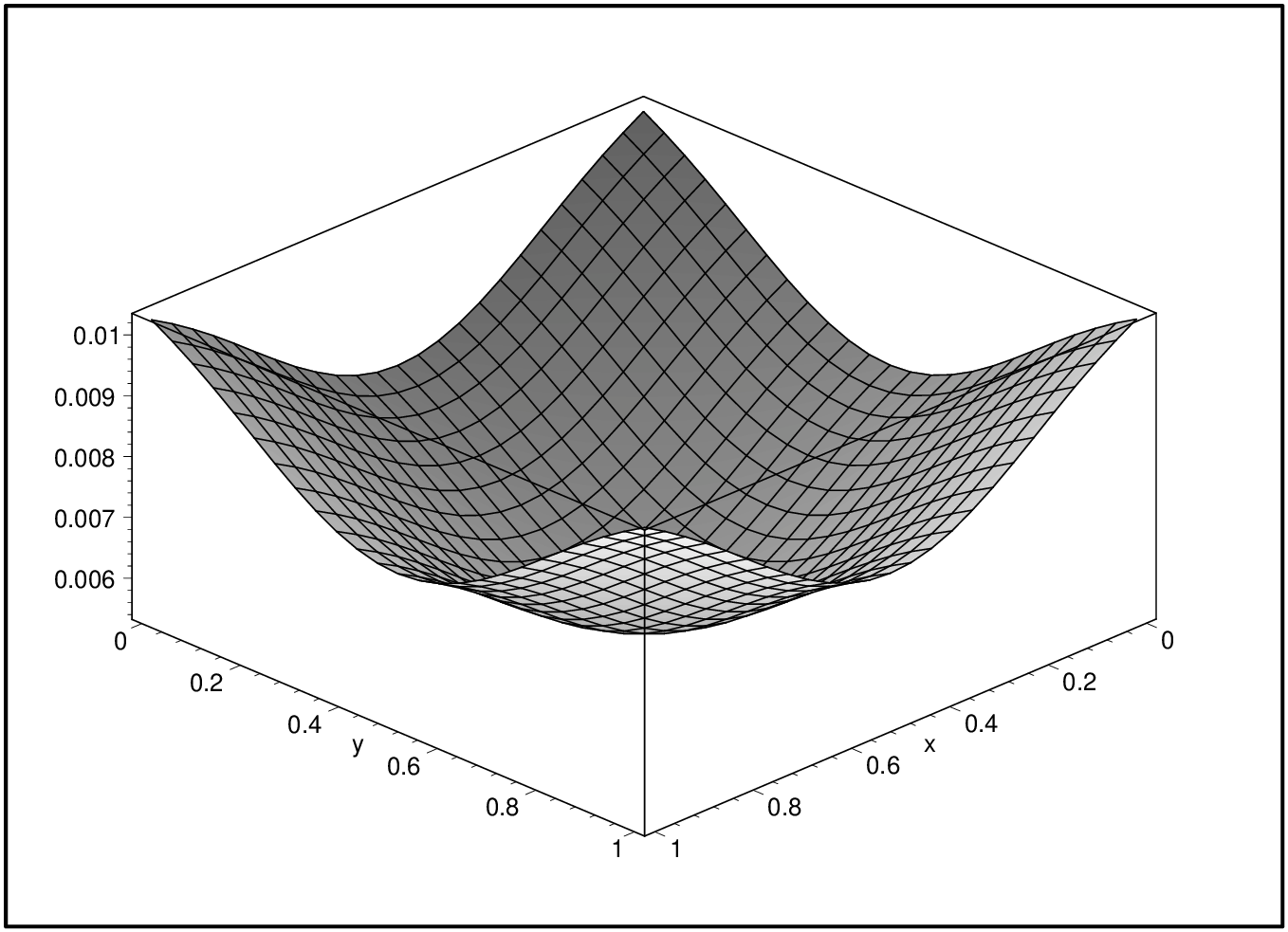, angle=270, width=11cm} }
\caption{Renormalized local energy density of the conformally coupled scalar field in the interior of the square waveguide.}
\label{Teta00S}
\end{figure}

\begin{figure}
\centerline {\epsfig{figure=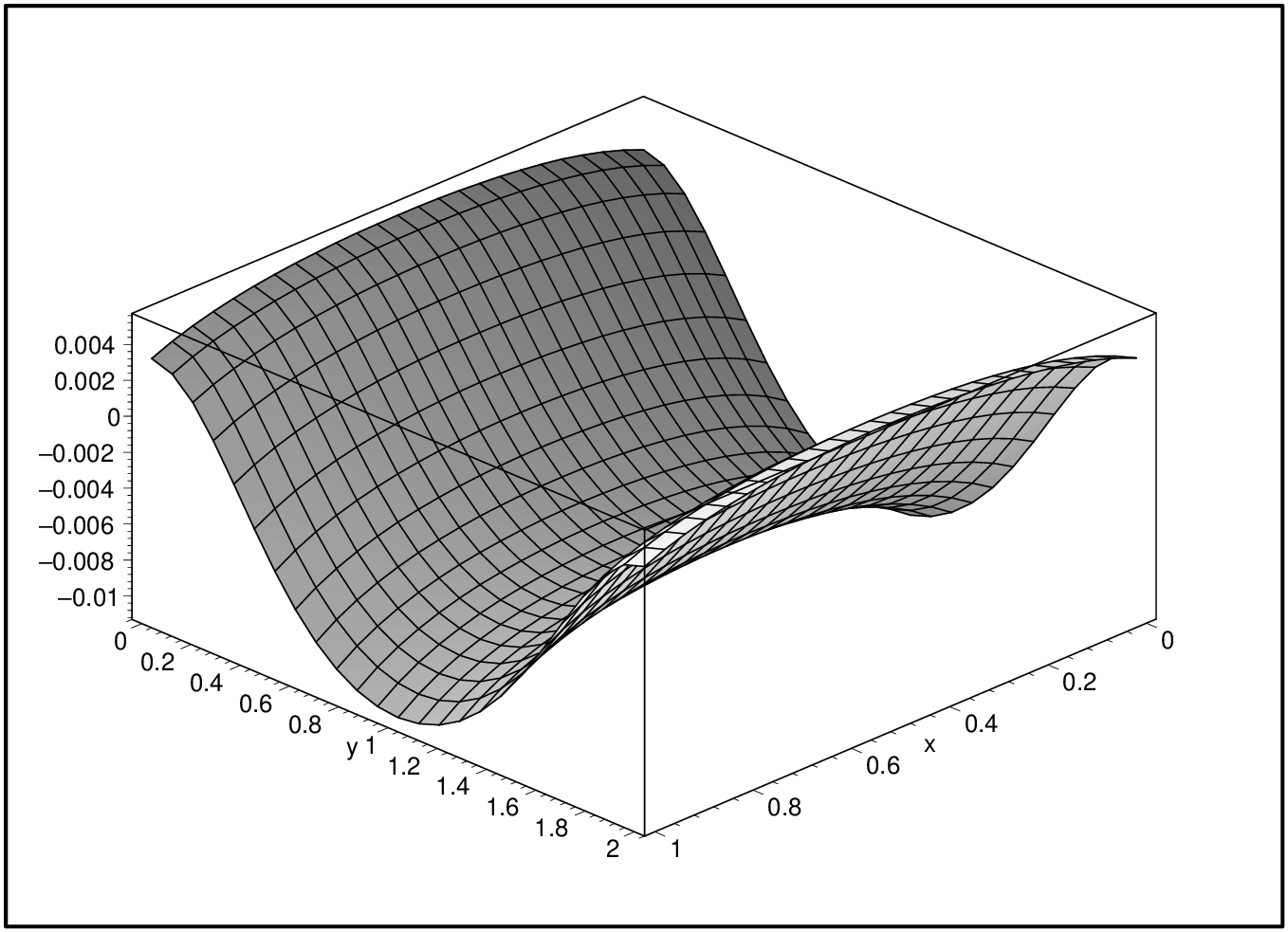, angle=270, width=11cm} }
\caption{Renormalized local energy density of the minimally coupled scalar field for the $b=2a$ waveguide.}
\label{T00R}
\end{figure}

\begin{figure}
\centerline {\epsfig{figure=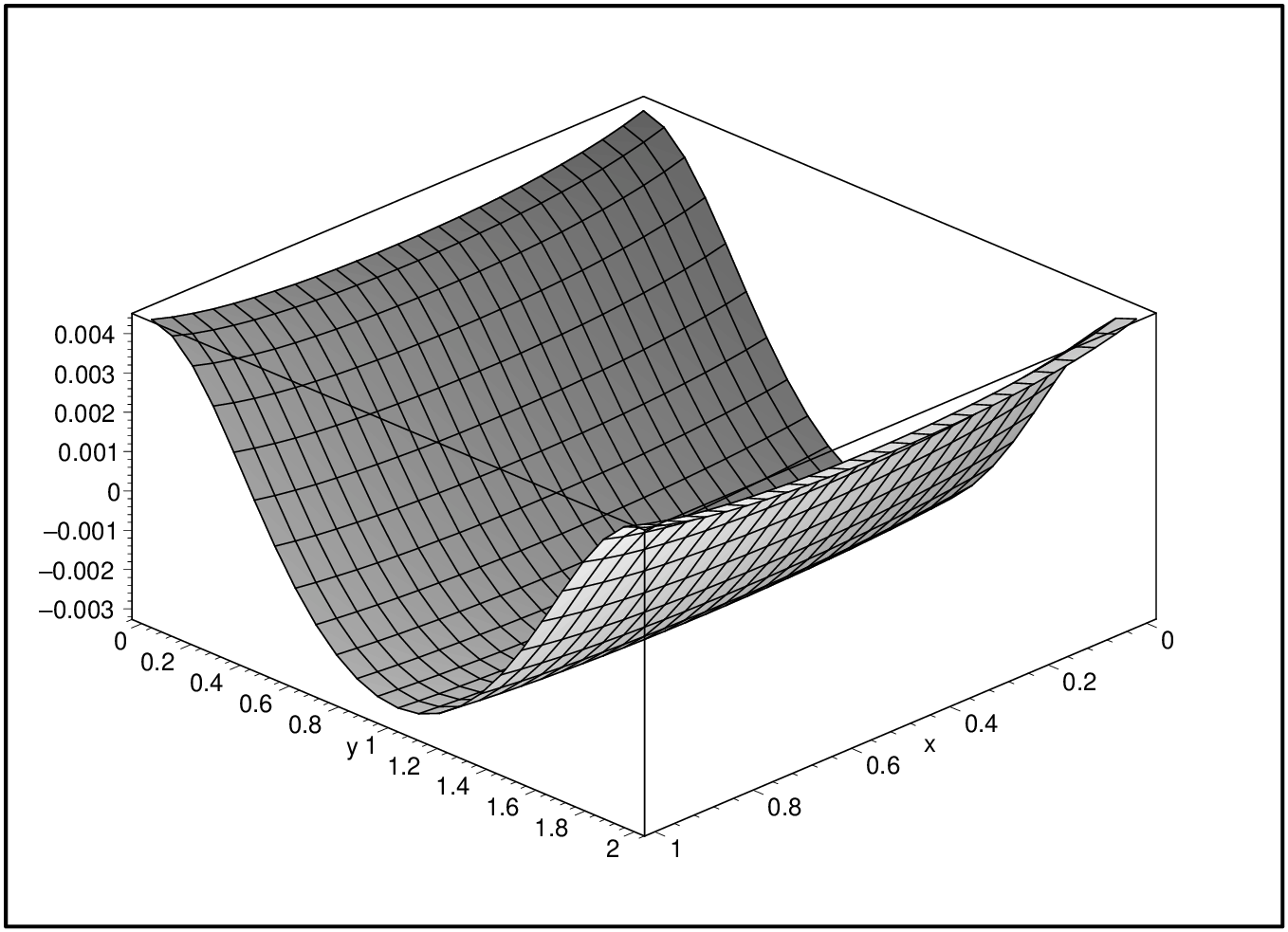, angle=270, width=11cm} }
\caption{Renormalized local energy density of the conformally coupled scalar field for the $b=2a$ waveguide.}
\label{Teta00R}
\end{figure}

An important lesson that we learn from eq.(\ref{LocGlob}) is that the integral
inside the waveguide of the finite part of $\left<T_{00}(\vec{x})\right>$
does not yield directly the total energy $E_C(a,b)$, but this one plus the
constant:
\begin{eqnarray}
 C(a,b)&=&\frac1{ab}\int_{ext}\left<T_{00}(\vec{x})\right>_B = \nonumber \\
 &&=-\frac1{24\pi^2}\left[\frac1{a^4} + \frac1{b^4} - \frac3{4a^2b^2}
    - \frac{3\arctan (b/a)}{4a^3b} - \frac{3\arctan (a/b)}{4ab^3}\right].
\end{eqnarray}
Dowker and Kennedy \cite{dow} have evaluated the total energy of the
conformally coupled scalar field in the interior of the waveguide for two
special configurations. For the square $a=b$, they showed that it assumes a
positive value. When $b=2a$ the energy decreases, assuming a negative value.
Figures (\ref{T00S}) and (\ref{Teta00S})\ show the form of $\left\langle
T_{00}(\vec{x})\right\rangle _{F}$ and $\left\langle \Theta _{00}(\vec{x}
)\right\rangle _{F}$ for the square waveguide, assuming $a=b=1$. They
present a minimal value in the middle of the waveguide and
assume only positive values, which produces a positive value. From the integral
of this density one should subtract the constant $C(1,1)$ in order to obtain
the total Casimir energy per unit area of the square waveguide. As the
value of $b$ increases (for $a=1$), these local quantities acquire negative
values in some space points, making the total energy decrease. Figures (\ref
{T00R}) and (\ref{Teta00R}) show the local energy density in the case $b=2a$. 
In this case the contribution of the negative part of the local energy
dominates and since one still has to subtract $C(1,2)$ from the integral of
this energy density, one obtains a negative total energy per unit area.

\section{Local forces}

In this section we will calculate the local Casimir force density that acts on
the walls of the waveguide. To do this we will use the relation between the
local force density and the discontinuity of the stress tensor across the
walls. Although we don't know the modes outside the waveguide (because the
external mode problem for the waveguide is unsolved), we can introduce an
external structure where the modes of the field are known \cite{bender}, in
such a way that the interior region is the interior of the waveguide. One way 
to do this is connecting two parallel infinite Dirichlet planes by two strips. 
In this configuration, we know the modes in all regions and the stress tensor 
can be calculated anywhere. Let us position two parallel infinite Dirichlet 
planes at $x_{1}=0$ and $x_{1}=a$ and connect these planes by two strips, 
positioned at $x_{2}=0$ and $x_{2}=b$. The interior region of this 
configuration is just the waveguide. In the regions $x_{1}>a$ and $x_{1}<0$ 
there are no contributions from the stress tensor to forces that act in the 
two infinite planes ($\left<T_{11}(\vec{x})\right>_{ext}=0 $). In the regions 
$0<x_{1}<a,\ x_{2}>b$ and $0<x_{1}<a,\ x_{2}<0,$ the components of the stress
tensor have a nonzero contribution to the forces that act on the strips. In
these regions the stress tensor has already been calculated in \cite{bender}.
For completeness, we present the relevant component here, i.e., 
$\left< T_{22}\left( \vec{x}\right) \right>_{ext}$: 
\begin{eqnarray}
 \left< T_{22}\left( \vec{x}\right) \right>_{ext}&=&\frac1{32\pi^2}
 \sum_{n=-\infty}^{\infty} \left((na)^{-4} + 2(na+x_1)^{-4}\right. \nonumber \\
 && \hspace{1cm}
 \left. -3\left[ (na+x_1)^2 + x_2^2 \right]^{-2} +
 4x_2^{-4} \left[1 + \left(\frac{na+x_1}{x_2}\right)^2 \right]^{3} \right).
\label{T22ext}
\end{eqnarray}
The equation above will serve to compute the local Casimir force that acts
on the strip at $x_2=0$ (a similar one exists for the strip at $x_2=b$).
We note that the edge divergences above at $(x_1,x_2)=(0,0),(a,0)$ will not be canceled, when we come to calculate the local force, by those of the interior 
of the waveguide that appear in eq.(\ref{t22}). Nevertheless neither the 
equation above nor eq.(\ref{t22}) present wall divergences as 
$x_2\rightarrow 0$. Thus the local Casimir force at the strip at $x_2=0$
diverges only at the edges, but not on the strip.

We note also that the components $\left< T_{21}\left( \vec{x}\right)
\right>_{ext} $ and $\left< T_{12}\left( \vec{x}\right) \right>_{ext} $
vanish on the walls. To obtain the local forces, we use the local force
density that acts on the point $\vec{x}$ and is given by $f_{i}\left( \vec{x}
\right) =-\partial _{j}T_{ij}\left( \vec{x}\right) .$ Thus the local force
per unit area on the boundary plane at $x_{1}=0$ is:
\begin{eqnarray}
\frac{F(x_{2})}{A} &=&\lim_{\varepsilon \rightarrow 0}
\left[ \left<T_{11}\left(x_{1}=-\varepsilon \right)\right> -
\left<T_{11}\left( x_{1}=\varepsilon \right)\right> \right] 
\nonumber \\
&=&\frac{1}{32\pi ^{2}}\sum_{n_{1},n_{2}=-\infty }^{\infty }\left(\right.
4\left[
(n_{1}a)^{2}+(n_{2}b)^{2}\right] ^{-3}\left[ n_{1}a\right] ^{2}+  \nonumber
\\
&&\hspace{1cm}
-4\left[ (n_{1}a)^{2}+\left( n_{2}b+x_{2}\right) ^{2}\right] ^{-3}\left[
n_{1}a\right] ^{2}+  \nonumber \\
&&\hspace{1cm}
-\left[ (n_{1}a)^{2}+(n_{2}b)^{2}\right] ^{-2}+\left[ (n_{1}a)^{2}+\left(
n_{2}b+x_{2}\right) ^{2}\right] ^{-2}\left.\right),  
\label{FemX2=0}
\end{eqnarray}
in the positive $x_1-$direction, and an equal but opposite force acts in the
plate at $x_1=a$. (We have to exclude the term $(n_1=n_2=0)$ in the third and 
fourth sums and the term $(n_1=0,n_2=-1)$ in the fourth sum. The last two 
exclusions accounts for the renormalization of the edge divergences.)

The force on the wall parallel to the plane at $x_{2}=0$ is given by 
\begin{eqnarray}
 \frac{F(x_{1})}{A} &=&\lim_{\varepsilon \rightarrow 0}\left[ \left\langle
 T_{22}\left( x_{2}=-\varepsilon \right) \right\rangle -\left\langle
 T_{22}\left( x_{2}=\varepsilon \right) \right\rangle \right]   \nonumber \\
&=&\frac{1}{32\pi ^{2}}\sum_{n=-\infty }^{\infty }\{\left( an\right)^{-4}
 -\left( an+x_{1}\right) ^{-4}\}+  \nonumber \\
&&-\frac{1}{32\pi ^{2}}\sum_{n_{1},n_{2}=-\infty }^{\infty }\left(\right.
 -4\left[
 (n_{1}a)^{2}+(n_{2}b)^{2}\right] ^{-3}\left[ n_{2}b\right] ^{2}+  \nonumber\\
&&\hspace{1cm}
 +4\left[ \left( n_{1}a+x_{1}\right) ^{2}+(n_{2}b)^{2}\right] ^{-3}\left[
 n_{2}b\right] ^{2}+  \nonumber \\
&&\hspace{1cm}
+\left[ (n_{1}a)^{2}+(n_{2}b)^{2}\right] ^{-2}-\left[ \left(
 n_{1}a+x_{1}\right) ^{2}+\left( n_{2}b\right) ^{2}\right] ^{-2}\left.\right)
 \nonumber\\
&=&\frac{1}{32\pi ^{2}}\sum_{n_{1},n_{2}=-\infty }^{\infty }
 \left(4\left[
 (n_{1}a)^{2}+(n_{2}b)^{2}\right] ^{-3}\left[ n_{2}b\right] ^{2}  
 -4\left[ \left( n_{1}a+x_{1}\right) ^{2}+(n_{2}b)^{2}\right] ^{-3}\left[
 n_{2}b\right] ^{2}\right)  +  \nonumber \\
&&+\frac{1}{16\pi ^{2}}\sum_{n_{1}=-\infty }^{\infty }\sum_{n_{2}=1}^{\infty}
\left(-\left[ (n_{1}a)^{2}+(n_{2}b)^{2}\right] ^{-2}+\left[ \left(
 n_{1}a+x_{1}\right) ^{2}+\left( n_{2}b\right) ^{2}\right] ^{-2}\right).
\label{FemX1=0}
\end{eqnarray}
An equal but opposite force acts on the wall at $x_{2}=b.$ (The edge 
divergences of eq.(\ref{T22ext}) do not cancel those of eq.(\ref{t22}), as 
we have stressed; nevertheless these were discarded when calculating the local
force above, and thus they do not appear.)

\begin{figure}
\centerline {\epsfig{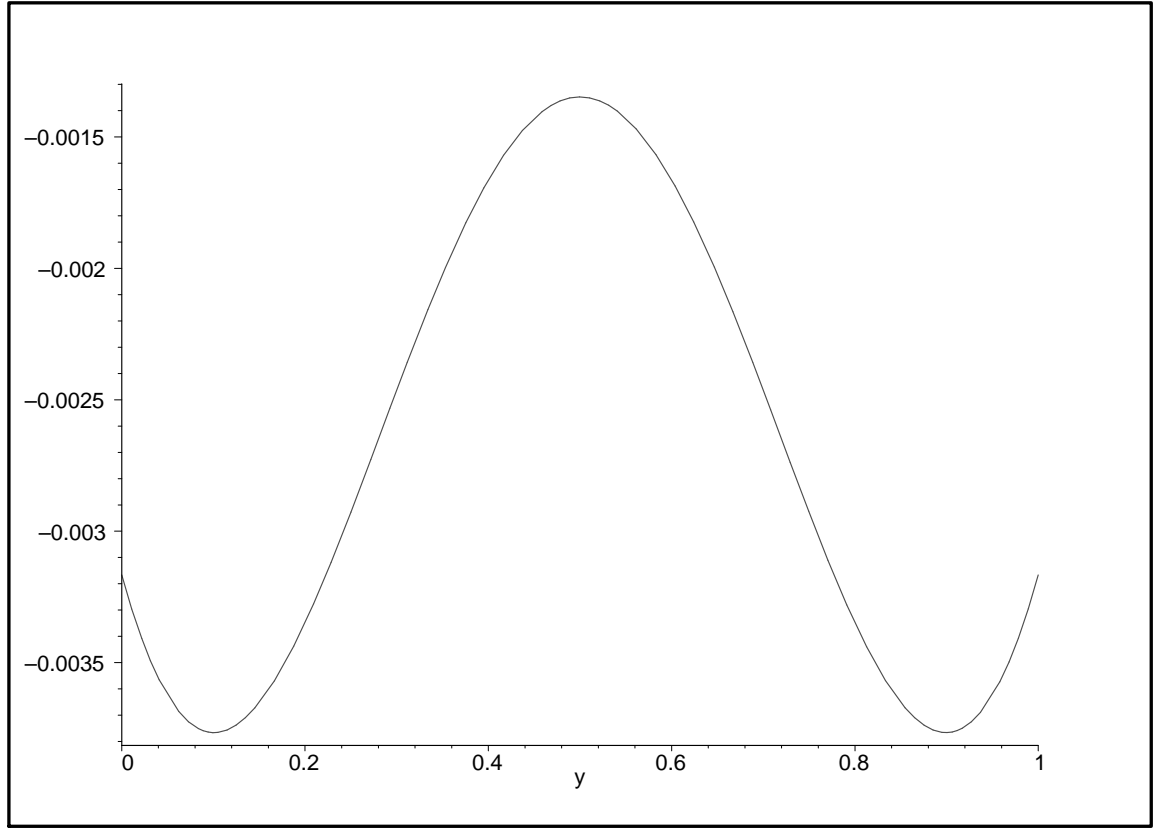} }
\caption{Renormalized local force density that acts on $x_{1}=0$ wall for the square waveguide.}
\label{Fx1S}
\end{figure}

\begin{figure}
\centerline {\epsfig{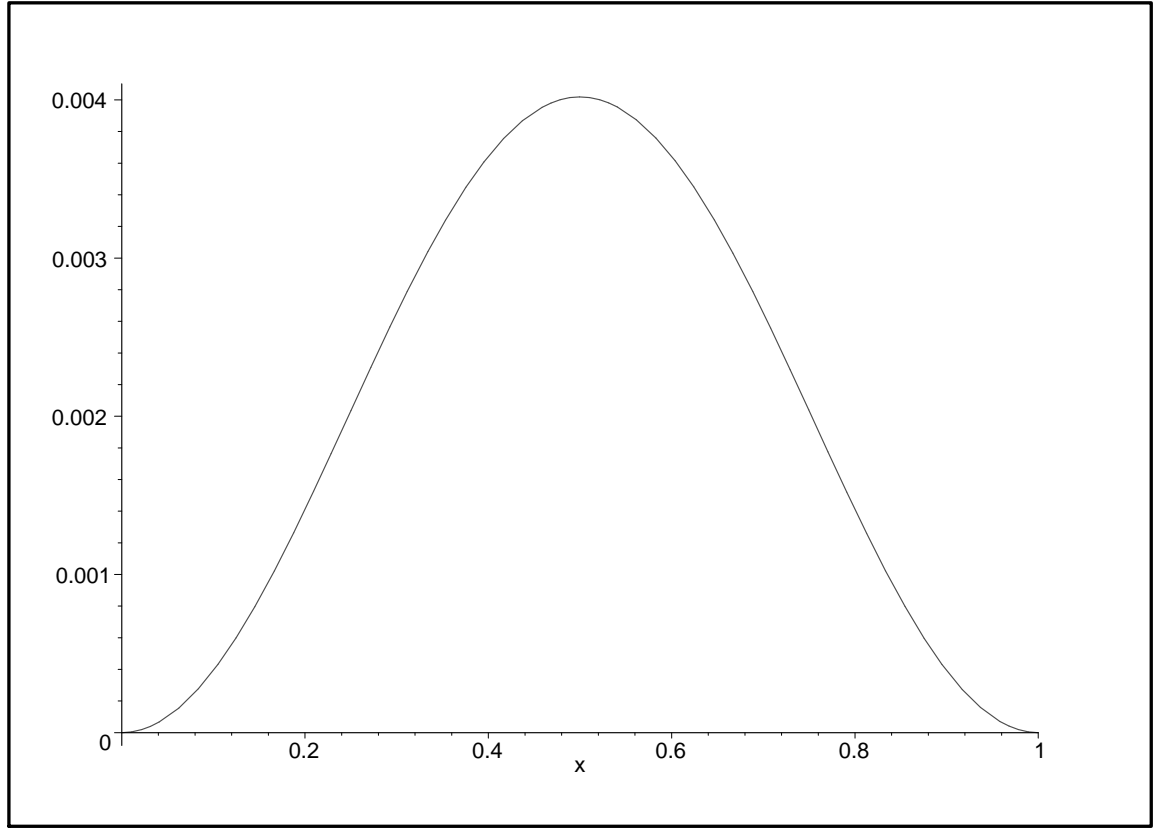} }
\caption{Local force density that acts on $x_{2}=0$ wall for the square waveguide.}
\label{Fx2S}
\end{figure}

Let us analyse how the local forces calculated above depend on the relative
sizes of the waveguide. Figure (\ref{Fx1S}) shows the dependence on $x_{2}$
of the finite part of the local force that acts on the wall parallel to the
plane $x_{1}=0$. It assumes only negative values and thus it is a repulsive force, in agreement with global calculations. The modulus of the force has a minimum in the middle of the wall and two maxima near the edges. Figure (\ref{Fx2S}) shows the depence on $x_{1}$ of the force on the wall parallel to the plane $x_{2}=0$. It is an attractive force but with only one maximum in the middle of the wall.
Although the global computation for the square waveguide gives a repulsive force in all walls, our attractive result is due to the external structure.

\begin{figure}
\centerline {\epsfig{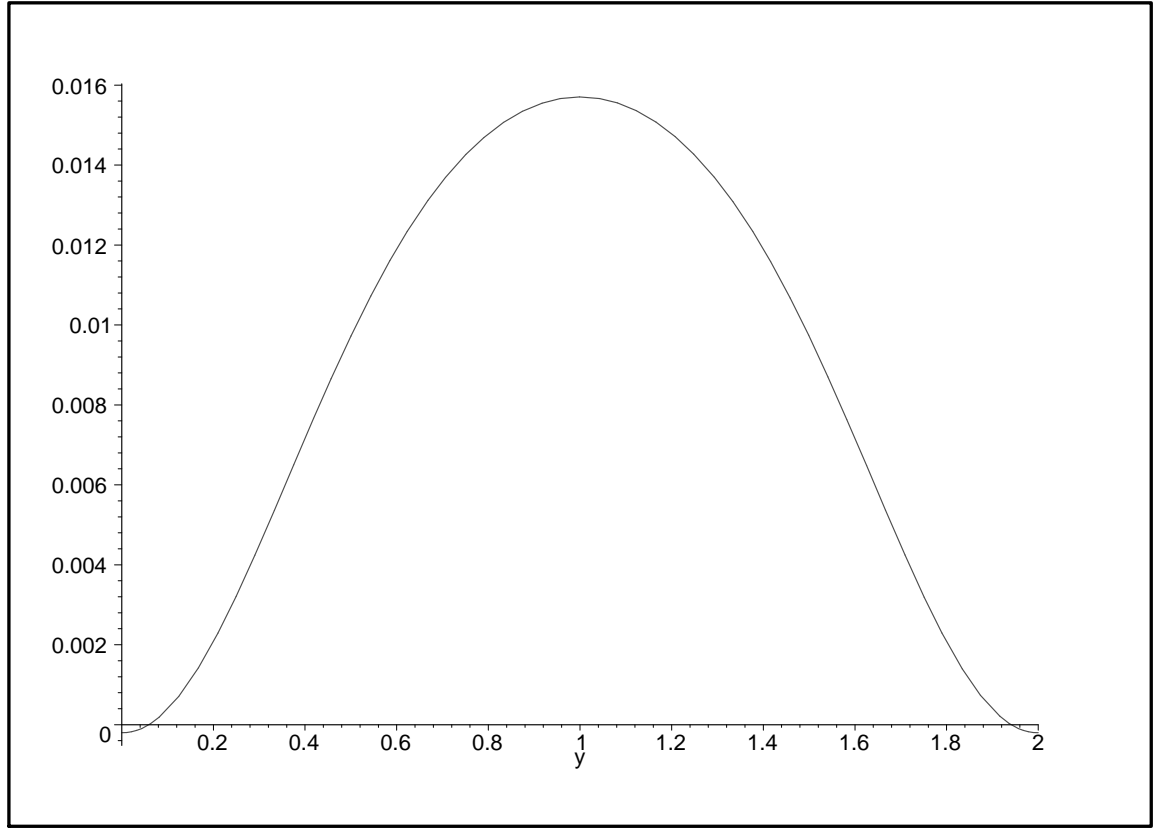} }
\caption{Renormalized local force density that acts on $x_{1}=0$ wall for the $b=2a$ waveguide.}
\label{Fx1R}
\end{figure}

\begin{figure}
\centerline {\epsfig{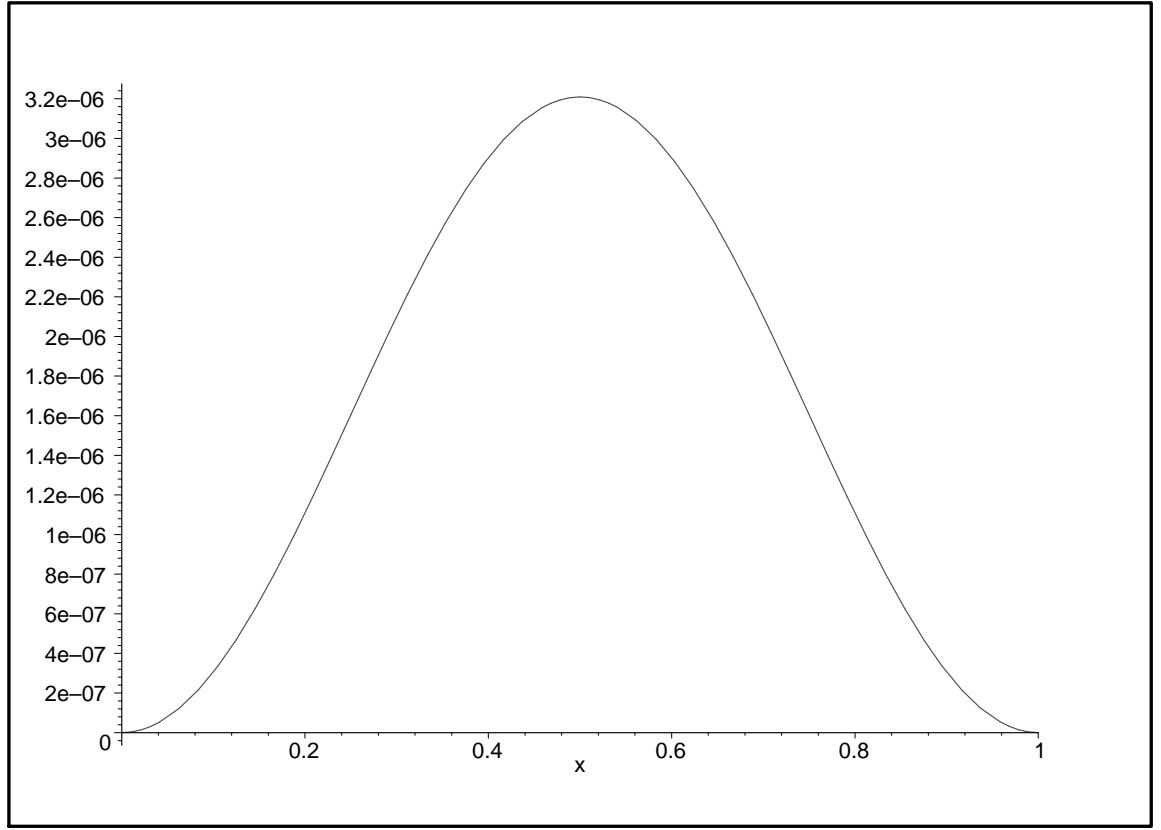} }
\caption{Local force density that acts on $x_{2}=0$ wall for the $b=2a$ waveguide.}
\label{Fx2R}
\end{figure}

\begin{figure}
\centerline {\epsfig{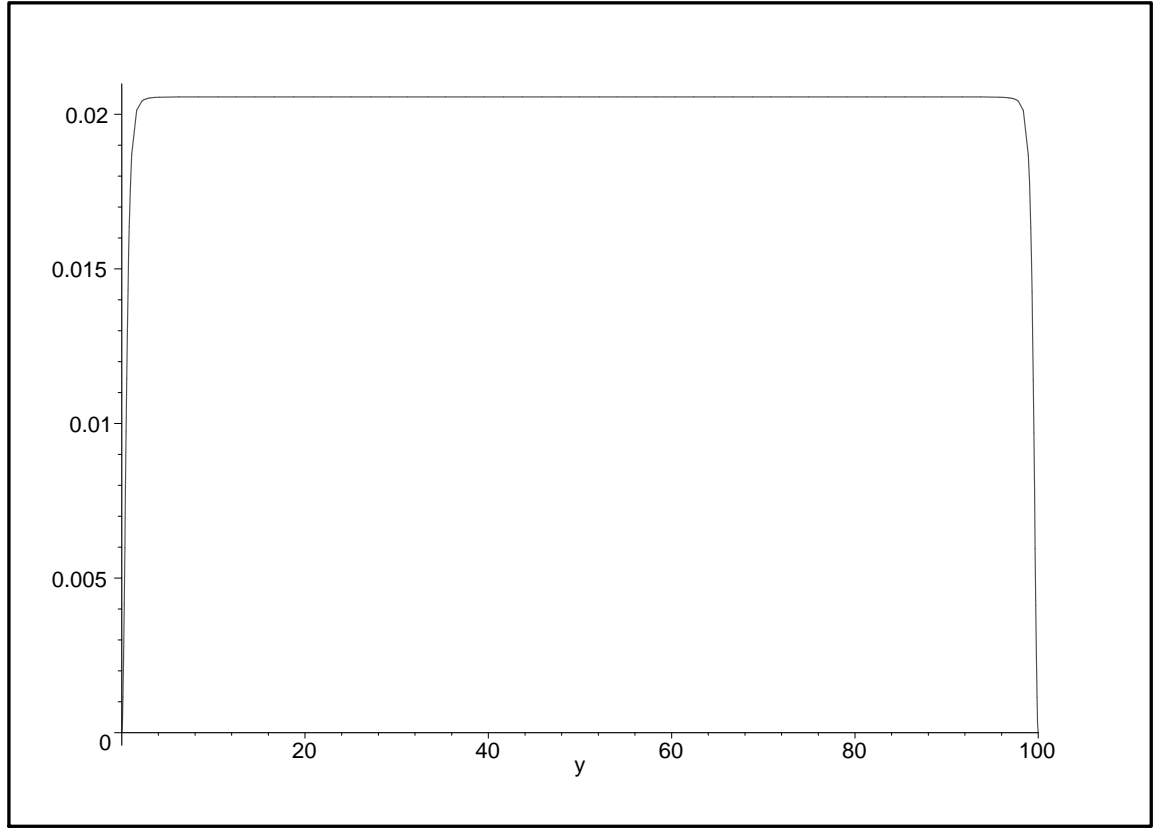} }
\caption{Renormalized uniform force density that acts on $x_{1}=0$ wall when $b>>a$.}
\label{Fx1U}
\end{figure}

Figures (\ref{Fx1R}) and (\ref{Fx2R}) show the forces that act on the walls
at $x_{1}=0$ and $x_{2}=0$ when $b=2a$. The local force at $x_{1}=0$ assumes
only positive values which makes it an attractive force, as we expect by
approaching the parallel plate configuration, but still highly non-uniform.
The force at $x_{2}=0$ assumes only positive values and it is very small in
comparison with the previous force. As $b$ grows, this force vanishes and
the force at $x_{1}=0$ behaves like the uniform Casimir force in the
parallel plate configuration as figure (\ref{Fx1U}) shows.

\section{Conclusions}

In this paper we obtained the canonical and the improved stress-energy
tensors of a massless scalar field in the interior of an infinitely long
waveguide. The result found is strongly position dependent as expected.
Although the global Casimir effect is related to experiments where we
measure the force between macroscopic surfaces, the local properties of the
vacuum field fluctuations can in principle be observed by measuring the
energy level shift of an atom interacting with the electromagnetic field. In
the case of the local problem, surface and edge divergences appear
related with the uncertainty principle. In order to compute the local forces
we introduced an external configuration for which it is possible to solve
the eigenmode problem. We have shown that the particular external configuration that we chose was not able to eliminate the wall and edge divergences of the interior of the waveguide. In order to eliminate them two possible ways are to 
take into account the real properties of the material, i.e., imperfect 
conductivity at high frequencies, or else make a quantum mechanical treatment 
of the boundary conditions, as was done by Ford and Svaiter \cite{ls}. An
alternative method of calculation (using a modified version of the Green's
function method) to find the renormalized stress-energy tensor associated
with the scalar field defined in the interior of an infinitely long
waveguide is under investigation by the authors.

We have also shown that the integral inside the cavity of the local result 
gives the known values for the global calculations, although the integral of
the {\it finite part} of $\left<T_{00}(\vec{x})\right>$ gives the total Casimir 
energy plus a constant dependent on the waveguide sizes $C(a,b)$.

\section{Acknowlegement}

This paper is supported by the Conselho Nacional de Desenvolvimento Cientifico 
e Tecnol\'{o}gico do Brasil (CNPq), and by Funda\c{c}\~ao de Amparo \`a 
Pesquisa do Estado de Minas Gerais (FAPEMIG).

\end{document}